\documentclass[conference]{IEEEtran}
\usepackage[letterpaper, left=1in, right=1in, bottom=1in, top=0.75in]{geometry}
\usepackage[utf8]{inputenc}
\usepackage[english]{babel}
\usepackage{amsmath}
\usepackage{amssymb}
\usepackage{bbm}
\usepackage{array}
\usepackage{cmap}
\usepackage{graphicx}
\usepackage{url}
\usepackage{cite}
\usepackage{hyperref}
\usepackage{siunitx}
\usepackage{tikz}
\usepackage{subfig}
\usepackage[figurename=Fig.]{caption}

\usetikzlibrary{arrows}
\usetikzlibrary{shapes.multipart}

\DeclareSIUnit{\dBm}{dBm}

\IEEEoverridecommandlockouts 
\newcommand\copyrighttext{%
\footnotesize \textcopyright \enspace 2017 IEEE. Personal use of this material is permitted. Permission from IEEE must be obtained for all other uses, in any current or future media, including reprinting/republishing this material for advertising or promotional purposes, creating new collective works, for resale or redistribution to servers or lists, or reuse of any copyrighted component of this work in other works. DOI: \href{https://doi.org/10.1109/PIMRC.2017.8292748}{10.1109/PIMRC.2017.8292748}
}
\newcommand\copyrightnotice{%
\begin{tikzpicture}[remember picture,overlay]
\node[anchor=south] at (current page.south) {\fbox{\parbox{\dimexpr\textwidth-\fboxsep-\fboxrule\relax}{\copyrighttext}}};
\end{tikzpicture}%
}

\begin{document}
\title{Mathematical Model of LoRaWAN \\ Channel Access with Capture Effect\thanks{The research was done at IITP RAS and was supported by the Russian Science Foundation (agreement No 14-50-00150).}}

\author{\IEEEauthorblockN{Dmitry Bankov\IEEEauthorrefmark{1}\IEEEauthorrefmark{2}, Evgeny Khorov\IEEEauthorrefmark{1}\IEEEauthorrefmark{3} and Andrey Lyakhov\IEEEauthorrefmark{1}\IEEEauthorrefmark{2}}
        \IEEEauthorblockN{\IEEEauthorrefmark{1}Institute for Information Transmission Problems, Russian Academy of Sciences, Moscow, Russia\\
	\IEEEauthorrefmark{2}Moscow Institute of Physics and Technology, Moscow, Russia\\
        \IEEEauthorrefmark{3}National Research University Higher School of Economics, Moscow, Russia\\
	Email: \{bankov, khorov, lyakhov\}@iitp.ru}
}

\maketitle
\copyrightnotice

\begin{abstract}
LoRaWAN is a promising low power long range wireless communications technology for the Internet of Things.
An important feature of LoRaWAN gateways is related to so-called capture effect: under some conditions the gateway may correctly receive a frame even if it overlaps with other ones.
In this paper, we develop a pioneering mathematical model of a LoRaWAN network which allows finding network capacity and transmission reliability taking into account the capture effect.  
\end{abstract}


\section{Introduction}
Low Power Wide Area Networks (LPWANs) are a promising solution for the Internet of Things, especially those related to sensor and actuator networks.
Recently LoRa/LoRaWAN has become a very popular LPWAN technology, mostly thanks to its reliable PHY design which enables energy-efficient long range communications and resilience to noise.
At the same time, LoRaWAN MAC protocol has fallen a victim of oversimplification due to the pursuit of keeping end devices cheap and easy-to-implement and therefore has many open issues \cite{bankov2016limits}.
Being an open solution operating in the ISM band, LoRaWAN faces an outlook of multiple independent networks using the same frequency range in vicinity of each other, which will result in problems related to interference.
Such a prospect could be avoided by the usage of an efficient power allocation and channel access scheme, however, its design requires an accurate model of network operation.

Although LoRaWAN specification was released in 2015, LoRa/LoRaWAN has already been extensively studied.
Most papers pay attention to LoRa PHY \cite{centenaro2016long, augustin2016study, haxhibeqiri2017lora}, leaving MAC out of scope.
Some researches have been directed towards performance evaluation of LoRaWAN networks in order to find the limits of this technology via simulation \cite{mikhaylov2016analysis, magrin2017performance}.
At the same time, little attention has been paid to analytical models.
Existing works \cite{adelantado2017understanding, reynders2017power} consider only unacknowledged transmission mode and Poisson total network traffic and apply the well-known approach for ALOHA modeling \cite{aloha}.
However, as shown in \cite{bankov2017mathematical}, such an approach provides incorrect results for basic LoRaWAN operation with acknowledgements (ACKs) and retransmissions, therefore more advanced mathematical model is required to describe the network.

In this paper, we extend the approach introduced in \cite{bankov2017mathematical} for LoRaWAN network modelling by taking into account the propagation losses and capture effect: i.e. the possibility of a packet to be received even if it intersects in time with packets from other devices but its power is sufficiently high.
The proposed model can be used to optimize, design and evaluate the performance of LoRaWAN MAC layer solutions, related to channel access, power and spreading factor control \cite{reynders2017power}.

The rest of the paper is organized as follows.
In Section \ref{sec:description} we explain LoRaWAN channel access and its features important for modelling.
Section \ref{sec:scenario} contains the description of the modelled scenario.
In Section \ref{sec:model} we describe the mathematical model of LoRaWAN channel access with capture effect.
Section \ref{sec:results} provides numerical results which prove the accuracy of the model.
Conclusion is given in Section \ref{sec:conclusion}.

\section{LoRaWAN Channel Access Description}
\label{sec:description}
A typical LoRaWAN \cite{lorawan} network consists of a server, gateways (GWs) and end devices, called \emph{motes}.
GWs are connected to the server via IP network and to the motes via LoRa links and act as relays between them.

We consider class A devices, which operate in a way fruitful for sporadic uplink data transmission. 

\begin{figure}[!t]
	\centering
	\begin{tikzpicture}[scale=0.7]
	\draw [arrows={-triangle 45}] (0,1.5) -- (10.0,1.4);
	\node at (1.3,  2.4) {\small{Main channel}};
	\node at (1.8,  1.1) {\small{Downlink channel}};
	\draw [arrows={-triangle 45}] (0,0.8) -- (10.0,0.8);
	\node at (9.7,  1.8) {$t$};
	\node at (9.7,  0.4) {$t$};
	\draw [line width=0.5mm] (1, 1.5) rectangle (3.6, 2.1);
	\node [align=center] at  (2.3, 1.8) {\small{Mote: Data}};
	\draw [line width=0.5mm] (5.0, 1.5) rectangle (7.4, 2.1);
	\node [align=center] at  (6.2, 1.8) {\small{GW: ACK1}};
	\draw [line width=0.5mm] (7.0, 0.8) rectangle (9.4, 1.4);
	\node [align=center] at  (8.2, 1.1) {\small{GW: ACK2}};
	\draw (3.6, 0) -- (3.6, 2.8);
	\draw (5.0, 2.2) -- (5.0, 2.8);
	\draw (7.0, 0) -- (7.0, 1);
	\draw [arrows={triangle 45-triangle 45}] (3.6,2.5) -- (5.0,2.5);
	\draw [arrows={triangle 45-triangle 45}] (3.6,0.2) -- (7.0,0.2);
	\node at (4.3,  2.7) {$T_1$};
	\node at (5.5,  0.5) {$T_2$};
	\end{tikzpicture}
	\caption{LoRaWAN channel access}
	\label{fig:channel_access}
	\vspace{-2.0em}
\end{figure}
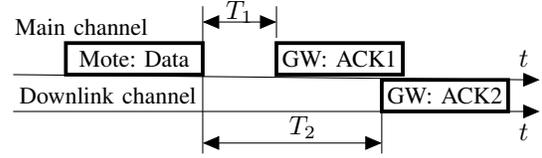

A LoRaWAN network simultaneously operates in several wireless channels.
For example, in Europe LoRaWAN devices can use three main channels and one downlink channel.
To transmit a data frame, each mote randomly selects one of the main channels (see Fig.~\ref{fig:channel_access}).
Having received the frame, the GW transmits two ACKs.
The first ACK is sent  $T_1$ after the frame reception in the same channel where data was transmitted.
The second one is sent after timeout $T_2 = T_1 + \SI{1}{\s}$ in the downlink channel.
If the mote does not receive any ACK, it makes a retransmission.
The standard recommends making a retransmission in a random time drawn from interval $[1, 1 + W]$ seconds, where $W = 2$.

At the PHY layer, LoRaWAN uses Chirp Spread Spectrum modulation.
Its main feature is that signals with different spreading factors can be distinguished and received simultaneously, even if they are transmitted in the same time on the same channel.
Channel width, coding rate and the spreading factor determine the data rate.
Lower data rates allow reliable transmission for big distance.
Data frames are sent at a rate determined by the GW, but the algorithm for rate allocation is not specified in the standard.
The first ACK is sent at a rate lower than the data rate for the frame transmission by a configurable offset (it can be zero).
The second ACK should always be sent at a fixed data rate, by default the lowest one.

\section{Problem Statement} 
\label{sec:scenario}

Consider a LoRaWAN network that consists of a GW and $N$ motes and operates in $F$ main channels and one downlink channel.
The motes use data rates $0, 1, ..., R$, set by the GW. 
The motes generate frames according to a Poisson process with total intensity $\lambda$ (the network load).
All motes transmit frames with 51-byte Frame Payload, which is the biggest payload that can fit a frame at the lowest data rate.
The frames are transmitted in the acknowledged mode, and ACKs carry no frame payload.
We consider a situation, when motes have no queue, i.e. if two messages are generated, a mote transmits only the most recent one.
Motes have a retry limit $RL$ and drop frames after $RL$ retransmission attempts.

We consider a scenario, when a frame is successful if for its duration its power is greater than the noise plus the power of the interfering frames transmitted in the same channel and at the same rate by at least $CR$ dB, where $CR$ is the co-channel rejection parameter, specified in LoRa chip datasheets.
For the described scenario, we want to find the maximal load at which the network can provide reliable communications.
We define packet error rate as the probability of frame transmission attempt to be unsuccessful and state the problem \emph{to find the packet error rate (PER) as a function of network load $\lambda$}.

\section{Mathematical Model}
\label{sec:model}

To solve the problem, we develop a mathematical model of the transmission process.
We consider separately the first and the subsequent transmission attempts.
As the first transmission attempts are described by the Poisson process, to find the PER in these assumptions, in Section \ref{first}, we consider the approach used to evaluate ALOHA networks\cite{aloha} and extend it to take into account ACKs.
This approach is however inapplicable for retransmissions, because they do not form a Poisson process, so in Section \ref{retries} we propose a way to take account of them.
For both kinds of attempts, we consider the possibility of several frames to be transmitted simultaneously and one of them to be successfully received if it has sufficient power.
Sections \ref{first} and \ref{retries} describe the general model of transmission process, while in Section \ref{special} we study a specific case, when the motes are distributed uniformly in a circle around the GW and the path-loss is described by Okumura-Hata model.

\subsection{The First Transmission Attempt}
\label{first}
Let $p_i$ be the probability of a mote using data rate $i$.
Its first transmission attempt is successful with probability:
\vspace{-0.7em}
\begin{equation}
\label{eq:success1}
P^{S,1}_i = P^{Data}_i P^{Ack}_{i},
\end{equation}
\vspace{-0.7em}

\noindent where $P^{Data}_i$ is the probability that the data frame is transmitted without collision at data rate $i$ and $P^{Ack}_i$ is the probability that at least one ACK is received by the mote, provided that the data frame is successful.

Since packets transmitted in different channels and at different rates do not collide, we need to consider separately each combination of channel and data rate.
The load at rate $i$ and one of $F$ channels equals $r_i = \frac{\lambda p_i}{F}$.

Probability of a successful data frame transmission consists of several summands.
Firstly, a data frame is successful if it does not intersect with another frame or an ACK of a previous frame.
Let $T^{Data}_{i}$ and $T^{Ack}_i$ be the durations of data frame and ACK, respectively, at rate $i$.
Intersection with a frame does not occur if no frames are generated in the interval $[-T^{Data}_{i}, T^{Data}_{i}]$, relative to the beginning of the considered frame.
For a Poisson process the probability of such event is $e^{-2 r_i T^{Data}_{i}}$. 
We consider that the GW cancels ACK transmission if it is receiving a data frame, so a collision with an ACK happens only if the ACK is generated in the interval $[-T^{Ack}_{i}, 0]$.
The rate of ACK generation is $P^{Data}_i r_i$, so the probability to avoid collision with an ACK is $e^{-r_i P^{Data}_i T^{Ack}_{i}}$.

Secondly, data frame is also successful, if it intersects with another frames, but the interfering signal is weaker.
The probability of $k$ motes transmitting their frames in time intervals that intersect the frame equals $e^{-2 r_i T^{Data}_{i}} \left(2 r_i T^{Data}_{i}\right)^k / k!$.
If we sum over all possible $k$ values, we obtain the equation for $P^{Data}_i$:
\begin{align}
\vspace{-1.0em}
P^{Data}_i = &e^{-(2 T^{Data}_{i} + P^{Data}_i T^{Ack}_{i}) r_i} + \nonumber\\
+ &\sum_{k = 1}^{N - 1} \frac{\left(2 r_i T^{Data}_{i}\right)^k}{k!} e^{-2 r_i T^{Data}_{i}} \mathbb{W}^{GW}_{i, k}, \label{eq:data1}
\vspace{-2.0em}
\end{align}
where $\mathbb{W}^{GW}_{i, k}$ is the probability of total interfering signal from $k$ motes (in dBm) having less power than the considered mote's signal plus co-channel rejection $CR$.
In this case, the signal power is measured at the GW.

The probability that at least one ACK arrives is calculated according to the inclusion-exclusion principle:
$P^{Ack}_i = P^{Ack1}_{i} + P^{Ack2}_i - P^{Ack1}_i P^{Ack2}_i$,
where $P^{Ack1}_{i}$ and $P^{Ack2}_i$ are the probabilities that the first and the second ACK, respectively, is transmitted successfully, provided that data was transmitted at rate $i$.

The first ACK is delivered if no data frame intersects it or if the interfering signal is weaker:
\begin{align}
\vspace{-0.4em}
P^{Ack1}_i =& e^{-\left(\min\left(T_1, T^{Data}_{i}\right) + T^{Ack}_{i}\right) r_i} + \nonumber\\
           +& \sum_{k = 1}^{N - 1} \frac{\left(r_i T^{Ack}_{i}\right)^k}{k!} e^{-r_i T^{Ack}_{i}} \mathbb{W}^{Mote}_{i, k},\label{eq:ack1}
\vspace{-0.4em}
\end{align} 
where in the first summand we take the minimum of $T^{Data}_{i}$ and $T_1$, because if a frame exceeds $T_1$, it breaks the acknowledged frame, but such an event is already taken into account by $P^{Data}_i$.
The second summand stands for the situation, when $k$ motes start transmission after the beginning of the ACK (if data is transmitted before ACK, the GW skips ACK transmission), but their total power is less than the power of the GW's signal, which happens with probability $\mathbb{W}^{Mote}_{i, k}$.
In this case, the signal power is measured at the ACK recipient mote, therefore the distributions of signal power in eqs. \eqref{eq:data1} and \eqref{eq:ack1} are different.

The second ACK is transmitted without collision if no data frame is successful in any other channel or at any other data rate, such that its second ACK would begin before the considered one and intersect it: 
\vspace{-0.4em}
\[
P^{Ack2}_i = e^{-T^{Ack}_{0} \lambda \left(1 - \frac{p_i P^{Data}_i}{F}\right) \sum_{j = 0}^{R} p_j P^{Data}_j},
\vspace{-0.4em}
\]
where in the exponent we multiply the interval during which the ACK should not start by the total intensity of successful frame process from all channels and all data rates except the one used by the considered mote.

\subsection{Retransmissions}
\label{retries}

The mote makes a retransmission when it does not receive any ACK after transmission
For that it randomly selects a channel and transmits data frame again after a random delay from $1$ to $1 + W$ seconds.

\begin{figure}[!t]
	\centering
	\begin{tikzpicture}[scale=0.7]
	\draw (0,1) -- (3.3,1);
	\draw [arrows={-triangle 45}] (4.1,1) -- (11,1);
	\node at ( 11,  0.6) {$t$};
	\node at (0.5,  0.6) {$0$};
	\node at (1.3,  0.6) {$x$};
	\node at (2.0,  0.6) {$T$};
	\node at (3.6,  1.0) {$...$};
	\node at (4.5,  0.6) {$\tau$};
	\node at (5.7,  0.6) {$y$};
	\node at (7.5,  0.6) {$z$};
	\node at (9.5,  0.6) {$\tau + W$};
	\draw [line width=0.3mm] (0.5, 1) rectangle (2.0, 1.5);
	\draw [dashed, line width=0.3mm] (1.3, 1) rectangle (2.8, 1.5);
	\draw [line width=0.3mm] (0.2, 2.1) rectangle (1.7, 2.6);
	\node at (3.7,  2.4) {frame of mote A};
	\draw [dashed, line width=0.3mm] (5.8, 2.1) rectangle (7.3, 2.6);
	\node at (9.3,  2.4) {frame of mote B};
	\draw [dashed] (4.5,  0.9) -- (4.5, 1.8);
	\draw [line width=0.3mm] (5.7, 1) rectangle (7.2, 1.5);
	\draw [dashed, line width=0.3mm] (7.5, 1) rectangle (9.0, 1.5);
	\draw [dashed] (9.5,  0.9) -- (9.5, 1.8);
	\end{tikzpicture}
\vspace{-0.5em}
	\caption{Retransmission}
	\label{fig:retransmission}
\vspace{-1.5em}
\end{figure}
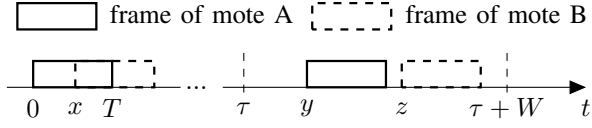

We consider a situation, when retransmission is caused by collision of two motes' (motes A and B) frames and find the probability of successful retransmission.
We use this probability as an amendment to the model to take account of the retransmissions.
Although it does not describe a case when more than two motes make a retransmission, it allows us to estimate $PER$ correctly, as shown in Section \ref{sec:results}.
There are two different cases.

The first case is when one frame is much more powerful than another ($w_A > w_B + CR$ or $w_B > w_A + CR$), so it is successful and the second one is needs a retransmission.
Let the probability of such event be $\mathbb{W}^{one}_i$.
In this case, retransmission is successful with the same probability as the first transmission attempt.

The second case is when both frames are unsuccessful, which happens with probability $\mathbb{W}^{Both}_i$.
In this case, both motes make a retransmission which is likely to result in a new collision, because frames in LoRaWAN are rather long compared with the window for random delay (compare $W = \SI{2}{\s}$ and $T^{Data}_0 \approx \SI{2.4}{\s}$).
Let 0 be the time when the frame of mote A begins, and $x$ be the offset for the frame of mote B (see Fig.~\ref{fig:retransmission}).
If motes choose different channels for retransmission, the collision is resolved.
Otherwise, with probability $\frac{1}{F}$, they choose the same channel.
In this case, let $y$ and $z$ be the times when motes A and B start their retransmission, respectively.
The value of $y$ is distributed uniformly in the interval $[\tau, \tau + W]$, where $\tau$ is the frame duration $T^{Data}_i$ plus the timeout for the ACK.
The value of $z$ is distributed uniformly in the interval $[\tau + x, \tau + x + W]$.
The retransmission results in a new collision, if one frame intersects with another or if a frame starts after the GW starts transmitting the acknowledgement for another frame.
Let $f\left(y, z, T^{Data}_i, T^{Ack}_i\right)$ be the indicator of such an event, i.e., it equals 1 if it happens and 0 otherwise:
\begin{align*}
f&\left(y, z, T^{Data}_i, T^{Ack}_i\right) = \mathbbm{1}\left(y                    \leq z \leq y + T^{Data}_i\right) +\\
       +& \mathbbm{1}\left(y + T^{Data}_i + T_1 \leq z \leq y + T^{Data}_i + T_1 + T^{Ack}_i\right)\\
       +& \mathbbm{1}\left(z                    \leq y \leq z + T^{Data}_i\right) +\\
       +& \mathbbm{1}\left(z + T^{Data}_i + T_1 \leq y \leq z + T^{Data}_i + T_1 + T^{Ack}_i\right).
\end{align*}

Using it we find the probability of a new collision:
\[
\vspace{-0.5em}
\resizebox{\linewidth}{!}{
$
P^c_i = \frac{1}{F} \cdot \frac{\int\limits_{-T}^{T} r_i e^{-r_i x} \int\limits_{0}^{W} \int\limits_{x}^{W + x} \frac{f\left(y, z, T^{Data}_i, T^{Ack}_i\right)}{W^2} dz dy dx}{\int\limits_{-T}^{T} r_i e^{-r_i x} dx}.
$
}
\]
Here we firstly integrate over all possible offsets between mote A and mote B frames that result in a collision, and then integrate over all possible random delays for both motes.
The integral in the denominator stands for the collision probability, so in the end we obtain the conditional probability of a new collision, provided that the first collision has already happened.

Taking into account all the aforementioned cases, the resulting probability of successful delivery of data frame during retransmission equals
\vspace{-0.4em}
\[ P^{Data}_{i, Re} = \frac{\left(\mathbb{W}^{One}_i + \mathbb{W}^{Both}_i \left(1 - P^c_i\right)\right)}{1 - \mathbb{W}^{GW}_{i, 1}} P^{Data}_i. \]
Here we divide the probabilities $\mathbb{W}^{One}_i$ and $\mathbb{W}^{Both}_i$ by $1 - \mathbb{W}^{GW}_{i, 1}$, because the fact that devices make a retransmission already means that the power of at least one mote is not enough for a successful transmission, and we thus obtain the conditional probabilities.

We obtain the probability of successful retransmission $P^{S, Re}_i$ as in eq. \eqref{eq:success1}, using $P^{Data}_{i, Re}$ instead of $P^{Data}_i$.
The probability of a successful transmission $P_{S}$ is
\[
P_{S} = \sum_i p_i \left(P_{1, i} P^{S, 1}_i + (1 - P_{1, i}) P^{S, Re}_i\right),
\]
where $P_{1}$ is the probability that the transmission attempt is the first one (not a retry).
$P_{1}$ is reverse to the average number of transmission attempts per a frame:
\vspace{-0.4em}
\[
\resizebox{\linewidth}{!}{
$P_{1, i} = \left(1 + \left(1 - P^{S, 1}_i\right) P^{G}_i \sum\limits_{r = 0}^{RL} \left(\left(1 - P^{S, Re}_i\right) P^{G}_i\right)^r \right)^{-1},$
}
\]
where $P^{G}_i$ is the probability of no new frame being generated during retransmission, which equals
\vspace{-0.4em}
\begin{align*}
P^{G}_i =& \frac{1}{W}\int\limits_{0}^{W} e^{-\frac{\lambda}{N}\left(T^{Data}_i + T_2 + T^{Ack}_0 + 1 + x\right)} dx = \\
= &\frac{N}{W \lambda} e^{-\frac{\lambda}{N}\left(T^{Data}_i + T_2 + T^{Ack}_0 + 1\right)} \left(1 - e^{-\frac{\lambda}{N} W}\right), 
\end{align*}
\vspace{-1.0em}

The packet error rate is calculated as $PER = 1 - P_{S}$.

The model estimates PER correctly up to such load, that new frames are generated so often that every retransmission results in a collision with a newly generated frame, which, in fact, is the network capacity.
To obtain it we divide $F$ by the average retransmission duration:
\vspace{-0.4em}
\[
\resizebox{\linewidth}{!}{
$\lambda^* = F \left(\sum_{i = 0}^{R} p_i \left( T^{Data}_{i} + T_2 + T^{Ack}_{0} + 1 + \frac{W}{2}\right) \right)^{-1}.$
}
\]

\subsection{Specific Case}
\label{special}

The general model contains values $\mathbb{W}^*_*$, to define which we consider a specific distribution of motes in space, a specific channel model and a way to assign data rates to the motes.
Let us consider the Okumura-Hata path-loss channel with no fading \cite{hata1980empirical}.
The signal power at the receiver equals $w_{rx}\left(d\right) = A - B \lg\left(d\right)$,
where $\lg(x)$ is base 10 logarithm of $x$,
$A = w_{tx} - 69.55 - 26.16 \lg\left(f\right) + 13.82 \lg\left(h_{GW}\right) + 3.2 \left(\lg\left(11.75 h_{Mote}\right)\right)^2 - 4.97$
and $B = 44.9 - 6.55 \lg\left(h_{GW}\right)$.
Here $w_{tx}$ is the transmit power (in dBM), $h_{GW}$ is the height of the GW antenna, $h_{Mote}$ is the height of the mote's antenna and $d$ is the distance between them.

\begin{figure*}[!t]
	\centering
	\subfloat[Data rate intervals and their bounds]{\includegraphics[width=1.7in]{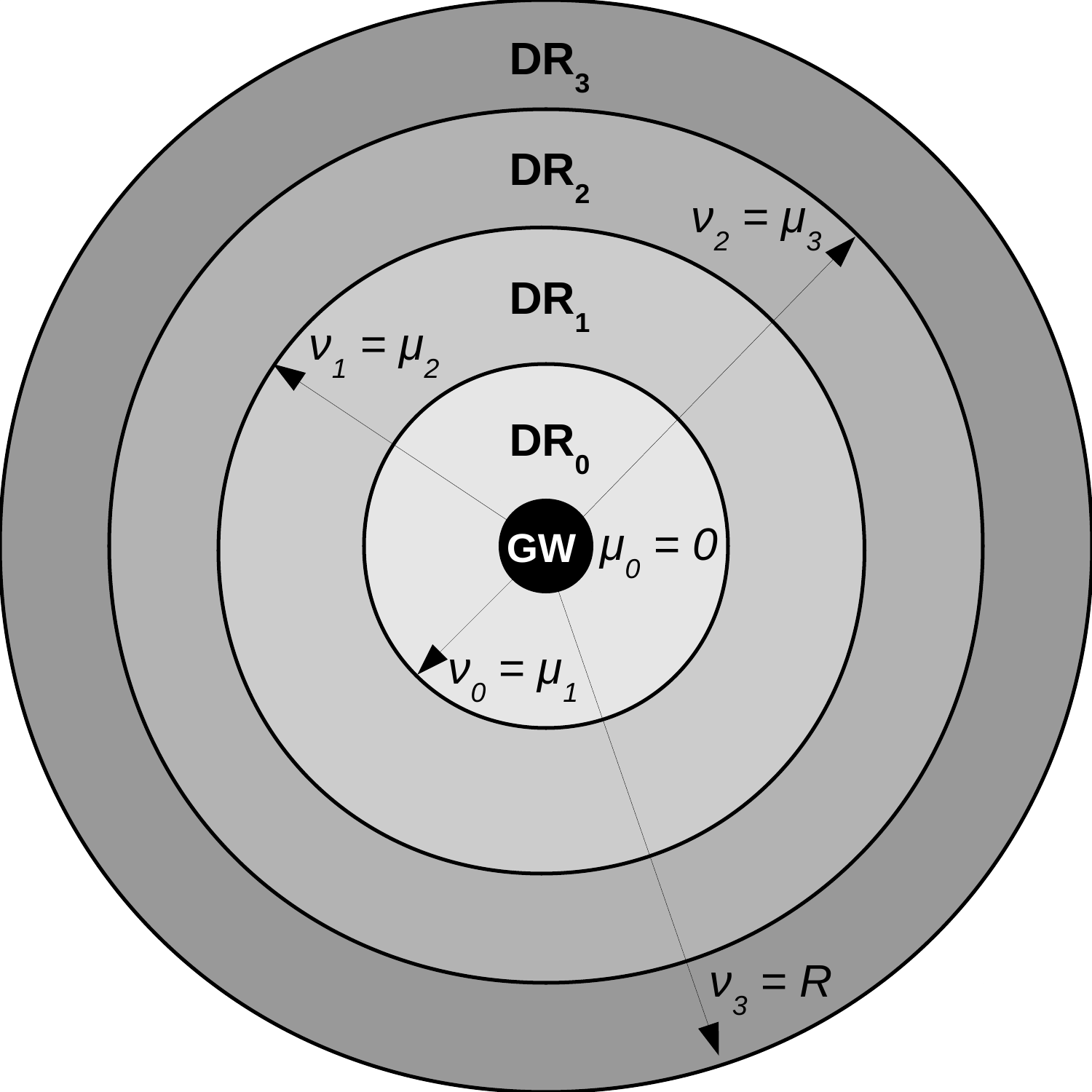}\label{fig:circles}}
	\hfil
	\subfloat[Mote 1 location for successful Mote 0 transmission]{\includegraphics[width=1.7in]{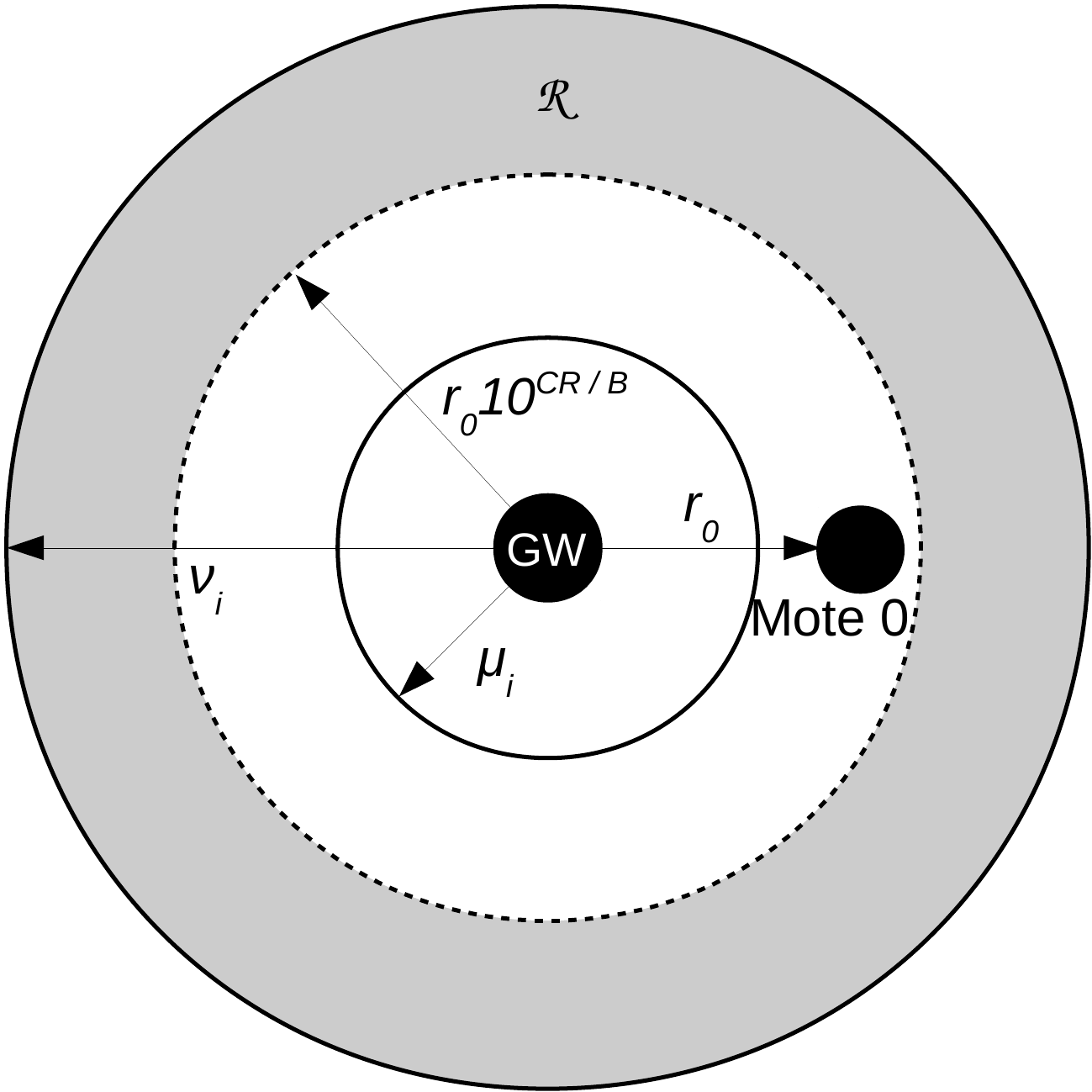}\label{fig:area1}}
	\hfil
	\subfloat[Mote 1 location when ACK is successful if it collides with data]{\includegraphics[width=2.2in]{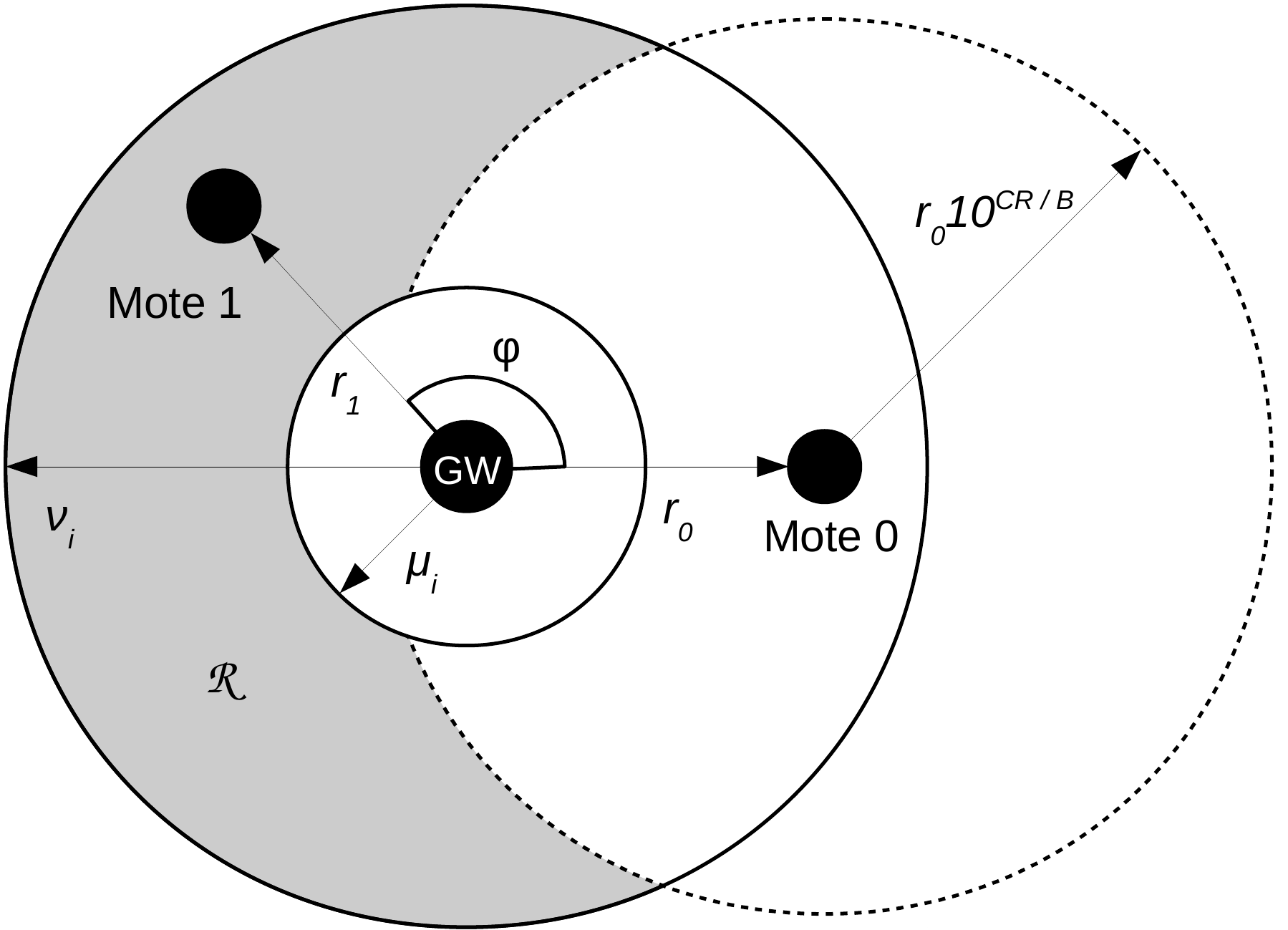}\label{fig:area3}}
	\caption{Mote locations for various cases.}
	\vspace{-1em}
\end{figure*}

Let the motes be spread around the GW uniformly in a circle with radius $R$.
In this case, the pdf of the Mote's distance from the GW equals $\rho(r) = \frac{2 r}{R}$.
We consider a case, when the server has a set of sensitivity thresholds $\left(w^{min}_i, w^{max}_i\right), w^{min}_{i + 1} = w^{max}_i$, and assigns data rate $i$ to a mote if the power of its signal at the GW is within $[w^{min}_i, w^{max}_i)$ interval.
These bounds define distance intervals $[\mu_i, \nu_i)$, within which motes use specific data rates (see Fig.~\ref{fig:circles}).
The radii are found as the solution of the equations $w_{rx}\left(\mu_i\right) = w^{min}_i$ and $w_{rx}\left(\nu_i\right) = w^{max}_i$.

Under such considerations we, firstly, derive the probability of a mote to use data rate $i$ as $p_i = \left(\nu_{i}^2 - \mu_{i}^2\right) / R^2$, which is the area, where the data rate is used divided by the total area of the circle.

Secondly, we consider the transmission of two motes, Mote 0 and Mote 1, and find the probability $\mathbb{W}^{GW}_{i, 1}$ of Mote 0 signal being more powerful than Mote 1 signal:
\vspace{-0.7em}
\begin{align*}
\mathbb{W}^{GW}_{i, 1} &= \mathbb{P} \left(w^{G}_1 < w^{G}_0 - CR\right) =\\
=& \mathbb{P} \left(r_1 > r_0 10^{\frac{CR}{B}}\right) = \int\limits_{\mu_i}^{\nu_i} \int\limits_{\mathcal{R}} \frac{4 r_0 r_1}{\left(\nu_{i}^2 - \mu_{i}^2\right)^2} d r_1 d r_0,
\vspace{-1.0em}
\end{align*}
where we integrate over distance $r_0$ from Mote 0 to the GW, and then integrate over such a distance $r_1 \in \mathcal{R}$ from Mote 1 to the GW that the power condition holds.
This location is shown in Fig. \ref{fig:area1} and is defined as
\[
\vspace{-0.5em}
\mathcal{R} = \left\{r_1: \mu_i <  r_1 < \nu_i \wedge r_0 \cdot 10^{\frac{CR}{B}} < r_1\right\}.
\]
The integral can be simplified as
\[
\vspace{-0.4em}
\mathbb{W}^{GW}_{i, 1} = \begin{cases}
\frac{\left(\nu_i^2 10^{-\frac{CR}{B}} - \mu_i^2 10^{\frac{CR}{B}}\right)^2}{2\left(\nu_i^2 - \mu_i^2\right)^2},	&	\nu_i > \mu_i 10^{\frac{CR}{B}},\\
0,															&	\nu_i \leq \mu_i 10^{\frac{CR}{B}}.
\end{cases}
\]

In a similar way, we find the probability $\mathbb{W}^{Both}_i$:
\[
\vspace{-0.4em}
\resizebox{\linewidth}{!}{
$
\mathbb{W}^{Both}_i = \begin{cases}
\frac{\nu_i^4 \left(1 - 10^{-\frac{2CR}{B}}\right) + \mu_i^4 \left(1 - 10^{\frac{2CR}{B}}\right)}{\left(\nu^2_{i} - \mu^2_{i}\right)^2},	&	\nu_i > \mu_i 10^{\frac{CR}{B}},\\
1,																		&	\nu_i \leq \mu_i 10^{\frac{CR}{B}}.
\end{cases}
$
}
\vspace{-0.4em}
\]

Then we find $\mathbb{W}^{One}_i = 1 - \mathbb{W}^{GW}_{i, 1} - \mathbb{W}^{Both}_i$.

Finally, we find the probability $\mathbb{W}^{Mote}_{i, 1}$ of the GW's signal at Mote 0 being more powerful than the signal from Mote 1:
\begin{align*}
\mathbb{W}^{Mote}_{i, 1} =& \mathbb{P} \left(w^{M}_1 < w^{M}_0 - CR\right) 
= \mathbb{P} \left(d_1 > r_0 10^{\frac{CR}{B}}\right)\\
=& \int\limits_{\mu_i}^{\nu_i} \iint\limits_{\mathcal{R}} \frac{2 r_0 r_1}{\left(\nu_{i}^2 - \mu_{i}^2\right)^2} \frac{d\phi}{\pi} d r_1 d r_0,
\end{align*}
where we firstly integrate over distance $r_0$ from Mote 0 to the GW, and then integrate over such distance $r_1$ and angle $\phi$ (see Fig. \ref{fig:area3}) that the power condition holds, i.e.:
\[
\resizebox{\linewidth}{!}{
$\mathcal{R} = \left\{r_1, \phi: \mu_i < r_1 < \nu_i \wedge cos \phi \leq \frac{r_0^2 + r_1^2 - r_0^2 10^{\frac{2CR}{B}}}{2 r_0 r_1}\right\}.$
}
\]

We further assume that $\mathbb{W}^{GW}_{i, k}, \mathbb{W}^{One}_i, \mathbb{W}^{Mote}_{i, k}$ equal zero for $k > 1$ (i.e., a device cannot retrieve a frame if it is interfered my two or more frames at once), which allows us to simplify the calculations of the model and, as numerical results show, does not affect the accuracy of the model.

\section{Numerical Results}
\label{sec:results}
\begin{figure}[tb]
	\center{\includegraphics[width=0.9\linewidth]{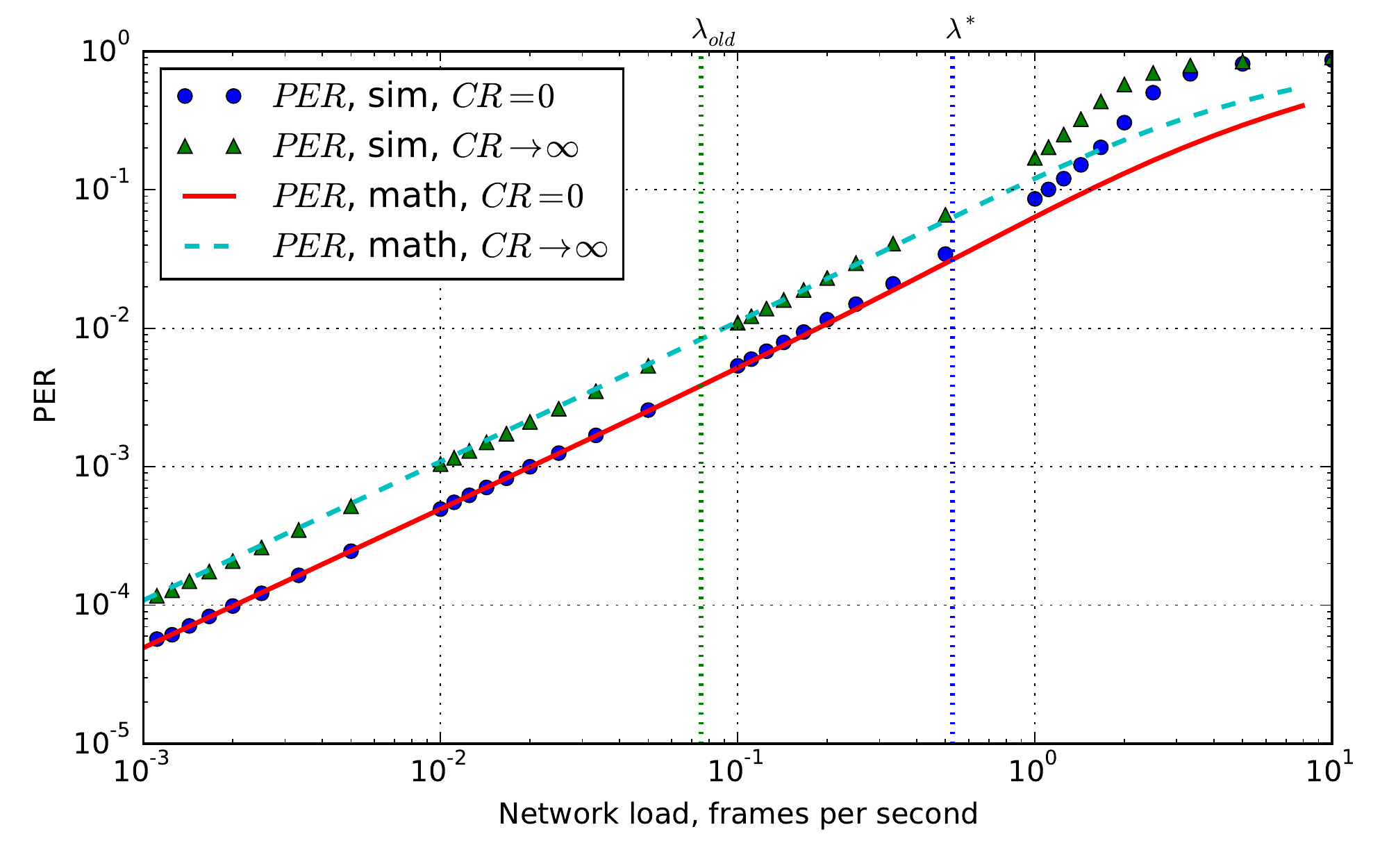}}
	\caption{Dependency of PER on the network load}
	\label{fig:per}
\vspace{-1.0em}
\end{figure}

To evaluate the performance of LoRaWAN networks, we consider EU 863-880 MHz ISM band with three default channels for data transmission and one downlink channel.
We simulate a network with 1000 motes and compare the average $\mathrm{PER}$ with one obtained with our mathematical model.
The results are shown in Fig.~\ref{fig:per}.
We show two opposite cases, the first one is $CR \rightarrow \infty$, which means that collision always happens if two frames intersect in time in one channel, and both frames are damaged in this case.
Such results completely align with the results of a model from \cite{bankov2017mathematical}, which does not take capture effect into account.
The second case is $CR$ = 0, which means that the frame always passes if its power is greater than interference plus thermal noise.
From Fig. \ref{fig:per} we can see that our model accurately predicts $PER$ for any $CR$ value.

According to numerical results, $PER$, calculated by taking into account the capture effect can be up to 50\% lower than the one calculated with traditional ALOHA approach.
As the result, the estimated maximal load supported by the network without exceeding the given $PER$, can be up to 100\% higher.

Numerical results also show that we correctly estimate $\lambda^*$, and that our upper bound on model accuracy is better than $\lambda^*_{old}$, which is the bound found in \cite{bankov2017mathematical}.  

\section{Conclusion}
\label{sec:conclusion}

In the paper, we consider LoRaWAN networks with class A devices operating in acknowledged mode.
We have extended the model from \cite{bankov2017mathematical} to accurately describe data transmission process, taking into account the difference between power of the signal from different devices and considering the capture effect.
We have developed a general model which can be used to evaluate performance of LoRaWAN networks in different scenarios, and also describe a specific case, when the propagation losses are described by Okumura-Hata model.
We provide the bound on network load for which our model provides correct results and show that the extended model is more accurate than the previous one.
This model can be used as a tool for future research related to power and spreading factor allocation in LoRaWAN networks, for parameter optimization and development of more advanced channel access schemes.

\bibliographystyle{IEEEtran}
\bibliography{biblio}

\begin{thebibliography}{10}
\providecommand{\url}[1]{#1}
\csname url@samestyle\endcsname
\providecommand{\newblock}{\relax}
\providecommand{\bibinfo}[2]{#2}
\providecommand{\BIBentrySTDinterwordspacing}{\spaceskip=0pt\relax}
\providecommand{\BIBentryALTinterwordstretchfactor}{4}
\providecommand{\BIBentryALTinterwordspacing}{\spaceskip=\fontdimen2\font plus
\BIBentryALTinterwordstretchfactor\fontdimen3\font minus
  \fontdimen4\font\relax}
\providecommand{\BIBforeignlanguage}[2]{{%
\expandafter\ifx\csname l@#1\endcsname\relax
\typeout{** WARNING: IEEEtran.bst: No hyphenation pattern has been}%
\typeout{** loaded for the language `#1'. Using the pattern for}%
\typeout{** the default language instead.}%
\else
\language=\csname l@#1\endcsname
\fi
#2}}
\providecommand{\BIBdecl}{\relax}
\BIBdecl

\bibitem{bankov2016limits}
D.~Bankov, E.~Khorov, and A.~Lyakhov, ``{On the Limits of LoRaWAN Channel
  Access},'' in \emph{Engineering and Telecommunication (EnT), International
  Conference on}.\hskip 1em plus 0.5em minus 0.4em\relax IEEE, 2016, pp.
  10--14.

\bibitem{centenaro2016long}
M.~Centenaro, L.~Vangelista, A.~Zanella, and M.~Zorzi, ``{Long-range
  Communications in Unlicensed Bands: The Rising Stars in the IoT and Smart
  City Scenarios},'' \emph{IEEE Wireless Communications}, vol.~23, no.~5, pp.
  60--67, 2016.

\bibitem{augustin2016study}
A.~Augustin, J.~Yi, T.~Clausen, and W.~M. Townsley, ``{A Study of LoRa: Long
  Range \& Low Power Networks for the Internet of Things},'' \emph{Sensors},
  vol.~16, no.~9, p. 1466, 2016.

\bibitem{haxhibeqiri2017lora}
J.~Haxhibeqiri, F.~Van~den Abeele, I.~Moerman, and J.~Hoebeke, ``Lora
  scalability: A simulation model based on interference measurements,''
  \emph{Sensors}, vol.~17, no.~6, p. 1193, 2017.

\bibitem{mikhaylov2016analysis}
K.~Mikhaylov, J.~Pet{\"a}j{\"a}j{\"a}rvi, and T.~H{\"a}nninen, ``{Analysis of
  the Capacity and Scalability of the LoRa Wide Area Network Technology},'' in
  \emph{{Proceedings of 22nd European Wireless Conference}}.\hskip 1em plus
  0.5em minus 0.4em\relax VDE, May 2016.

\bibitem{magrin2017performance}
D.~Magrin, M.~Centenaro, and L.~Vangelista, ``Performance evaluation of lora
  networks in a smart city scenario,'' in \emph{IEEE ICC 2017}, 2017.

\bibitem{adelantado2017understanding}
F.~Adelantado, X.~Vilajosana, P.~Tuset, B.~Martinez, J.~Melia-Segui, and
  T.~Watteyne, ``{Understanding the Limits of LoRaWAN},'' \emph{{IEEE
  Communications Magazine}}, Jun. 2017.

\bibitem{reynders2017power}
B.~Reynders, W.~Meert, and S.~Pollin, ``Power and spreading factor control in
  low power wide area networks,'' in \emph{IEEE ICC 2017}, 2017, pp. 1--5.

\bibitem{aloha}
N.~Abramson, ``{THE ALOHA SYSTEM: Another Alternative for Computer
  Communications},'' in \emph{Proceedings of the November 17-19, 1970, Fall
  Joint Computer Conference}, ser. AFIPS '70 (Fall).\hskip 1em plus 0.5em minus
  0.4em\relax New York, NY, USA: ACM, 1970, pp. 281--285.

\bibitem{bankov2017mathematical}
D.~Bankov, E.~Khorov, and A.~Lyakhov, ``{Mathematical Model of LoRaWAN Channel
  Access},'' in \emph{Proceedings of the IEEE WoWMoM, Macao, China}, 2017, pp.
  12--15.

\bibitem{lorawan}
\emph{{LoRaWAN Specification, V. 1.0.2}}, LoRa Alliance, Jul. 2016.

\bibitem{hata1980empirical}
M.~Hata, ``Empirical formula for propagation loss in land mobile radio
  services,'' \emph{IEEE transactions on Vehicular Technology}, vol.~29, no.~3,
  pp. 317--325, 1980.

\end{thebibliography}

\end{document}